\documentclass[conference]{IEEEtran}
\IEEEoverridecommandlockouts
\usepackage{cite}
\usepackage{amsmath,amssymb,amsfonts}
\usepackage{algorithmic}
\usepackage{graphicx}
\usepackage{textcomp}
\usepackage{xcolor}
\usepackage{hyperref}
\usepackage{booktabs}
\usepackage{orcidlink}
\hypersetup{pdfborder={0 0 0}}

\def\BibTeX{{\rm B\kern-.05em{\sc i\kern-.025em b}\kern-.08em
    T\kern-.1667em\lower.7ex\hbox{E}\kern-.125emX}}
    
\begin{document}

\title{Teaching an Online Multi-Institutional Research Level Software Engineering Course with Industry - an Experience Report}
\author{%
  Pankaj Jalote \orcidlink{0009-0001-8552-8394}\\IIIT Delhi, India\\jalote@iiitd.ac.in\\
  \and Y. Raghu Reddy \orcidlink{0000-0003-2280-5400}\\IIIT Hyderabad, India\\raghu.reddy@iiit.ac.in
  \and Vasudeva Varma  \orcidlink{0000-0003-1923-1725}\\IIIT Hyderabad, India\\ vv@iiit.ac.in
}

\maketitle
\begin{abstract}
Covid has made online teaching and learning acceptable and students, faculty, and industry professionals are all comfortable with this mode. This comfort can be leveraged to offer an online multi-institutional research-level course in an area where individual institutions may not have the requisite faculty to teach and/or  research students to enroll. If the subject is of interest to industry, online offering also allows industry experts to contribute and participate with ease. Advanced topics in Software Engineering are ideally suited for experimenting with this approach as industry, which is often looking to incorporate advances in software engineering in their practices, is likely to agree to contribute and participate. In this paper we describe an experiment in teaching a course titled “AI in Software Engineering” jointly between two institutions with active industry participation, and share our and student's experience. We believe this collaborative teaching approach can be used for offering research level courses in any applied area of computer science by institutions who are small and find it difficult to offer research level courses on their own. 

\end{abstract}

\begin{IEEEkeywords}
Collaborative teaching, Multi-institution teaching, Online teaching, Teaching advanced courses, Teaching Software Engineering, Research-level course.
\end{IEEEkeywords}

\section{Introduction}
Many smaller academic institutions (particularly in India) have modest faculty sizes and small PhD programs. As a result, As a result, they primarily offer core courses or broadly popular electives, struggling to support advanced, research-oriented electives. Consequently, these institutions are generally not able to offer research oriented courses on advanced topics which can help PhD students and Masters students researching those areas. For example, software engineering (SE) research courses are rarely available because only one or two faculty might specialize in SE as most of the SE faculty are still part of broader CS departments, and typical policies discourage running courses with very few enrollments.
 
In such a situation, a simple approach can be to pool resources, where faculty and students across multiple institutions are part of research-level courses in advanced topics within SE. Historically, inter-institutional courses were rare due to logistical barriers (e.g., travel between campuses), but the widespread adoption of online teaching post COVID-19 lowers these barriers. By leveraging online platforms, faculty from different institutions can co-teach a course, and even a handful of students per campus can together form a viable class. Also, if the topic is of interest to industry, working professionals can enroll online and industry experts can contribute lectures, enriching the course experience.

In this paper, we detail our experiences in teaching a course on “Artificial Intelligence (AI) in Software Engineering (SE)” between two academic institutions located in two different cities in India, with a strong participation from industry.  AI for SE has been an important area of research both in academia and industry, and hence well suited for the approach we wanted to try. The area has become extremely active since the advent of LLMs. This course was offered in 2021, before LLMs became popular, hence the contents of the course were different than what they potentially would be today. 

Generally, one of the key goals of a research-level course is to expose the students to advanced topics in the subject of interest. Such courses often involve students taking up some topic and writing a report on it and making a presentation to the class based on state of the art on that topic. A research-level course may also expect students to engage in some small research project. This course had all these as the objectives. The learning outcomes for the course were: (i) Gaining familiarity on AI methods/techniques being used for addressing different problems in SE, (ii) In-depth understanding of use of AI models in one particular SE issue, (iii) Improving ability of the participants to review research literature and understand the state of the art in an area, and (iv) Improving ability of the participants to do independent research.

In this experience report, we detail the challenges of delivering a multi-institution, research oriented software engineering course with industry involvement and how we addressed them.  The paper's contributions are: (1) an account of a novel multi-institution, industry-partnered course model in software engineering, (2) identification of key challenges in implementing such a course (with the solutions we employed, and (3) an evaluation of the outcomes.

\section{Related work}
There is a considerable body of work on software engineering education. Much of it focuses on basic concepts in software engineering and the pedagogy used to teach them. For example, there are reports on using game design to teach software engineering\cite{b12}, gamifying software engineering tasks \cite{b13}, aligning course contents with industry (including startups) \cite{b14}, using agile methods, and other processes \cite{b15}\cite{b16}.  There is also some work on teaching advanced courses that focus on specific topics like requirements engineering \cite{b17}, software architecture\cite{b18}, design \cite{b19}, quality engineering \cite{b20}\cite{b21}, code comprehension \cite{b22}, etc., and on teaching courses involving research activities \cite{b23}.

Courses on global or distributed software engineering education have been taught for many years (from before online teaching became widespread) and there are many reports on different aspects of it (e.g. \cite{n1, n2, n3}). Generally in these courses, the focus of collaboration is among students from different institutions in different countries/institutions with the main purpose of developing experience (and skills) in executing a (globally) distributed software. The instructors need to collaborate among themselves largely for ensuring that the project can be done successfully by students from different universities.   The collaborative teaching experiment we describe is quite different in goals as well as in execution. The focus is not on executing distributed projects by students, but on distributed teaching by combining strengths and knowledge of instructors in different institutions to offer a richer course for students of different institutions.

As many universities are becoming multi-campus universities, there are efforts to have single campus courses to be extended to multi campus setting \cite{n4, n5}. The general approach in these is to have one campus design the course which is then delivered in different campuses by their teachers, or deliver the lecture using technology in different campuses. This is very different from the goal of our experiment. Our goal is to leverage strengths of faculty from different institutions to offer a richer course for students. Also, these efforts are at institution level and require administrative support and regulations. The approach we describe is for faculty to collaborate in teaching a research level course, leveraging the existing rules and regulations, with minimal special support for collaboration from the institutions.

During COVID, there were also efforts to have seminar series for students in multiple institutions with lectures from experts from different institutions on some subjects. These were largely to provide and share knowledge, often not as a formal credit based course. Metchik et al (\cite{n6}) describe how they used this approach to tap the extra time available during covid to medical residents for additional training .  Collaborative online teaching with multiple instructors has also been tried in courses with multiple sections and multiple instructors \cite{n7}. The instructors in this case are in the same university and the students are also from that university. The main challenge addressed is how to use online mode for a multi-section classroom and do multiple projects.

Authors have published papers on collaborative teaching, also referred to as co-teaching. Veteska et al. \cite{b1} have done an exhaustive study on co-teaching in schools by exploring articles published in 16 different databases between the years 2005 and 2020. Co-teaching methods have been applied in liberal arts colleges for specific courses (e.g. English as a foreign language) with positive learning outcomes \cite{b2}\cite{b3}. In Software Engineering, project based learning courses, game based learning courses have been known to involve multiple faculty including industry faculty to deliver the course material \cite{b4}. Many courses use co-teaching and learning methods at an undergraduate level. They use free and open source software (FOSS) and some form of mentorship from industry to enhance their software product engineering skills \cite{b6}\cite{b5}. Frameworks have been proposed to improve collaborative teaching in online software engineering education \cite{b5}. 

To the best of our knowledge, there are no reports on multi-institution collaborative teaching of research-level courses where the cohort consists of faculty and students from multiple institutions and industry. We believe this approach of multiple institutions collaboratively teaching research-level courses with participation of industry is novel and interesting, and one which can be leveraged by smaller institutions across the globe to offer research level courses in any applied area of computer science. 

\section{Challenges in Multi Institutional Course Offering}

A course dealing with advanced research topics must be interactive and facilitate discussion. Hence it needs to be offered in real-time online mode, i.e. pre-taped lectures model is unsuitable as it allows for no real-time interaction and discussion. We faced multiple challenges and issues in actually offering such a course in a multi-institution mode.  We discuss some of the challenges and how we addressed them. 

\begin{itemize}
    \item \textbf{Selection of the topic}: The chosen topic must be an active research area of mutual interest that benefits all participating faculty and students. If industry professionals are to be involved as learners or guest speakers, the topic should also appeal to industry. We selected “AI in SE” as it met these criteria, being of interest in both research and industry.
    
    \item \textbf{Course Approval}: Without formal agreements (e.g., a Memorandum of Understanding), jointly offering a single course across institutions is difficult. The approach we took is that while the course may be co-taught by faculty from a few institutions, it is offered as an elective in each institution separately, being taught by faculty from that institution. For example, let us say faculty members A, B, and C from three institutions want to offer a course “Special topics in SE”. The course is floated independently in each of the institutions with the faculty of that institute being the instructor. The course is scheduled at the same time in all the institutes. The lectures in the course are divided between the instructors A, B, and C. Lectures delivered by faculty from the other institute were treated as “guest lectures” in each course offering. As guest lectures from qualified people are permitted in most institutions, this approach can work within the existing regulatory framework of most institutions.
   
    \item \textbf{Scheduling the course}: 
    Coordinating schedules across two universities was a key challenge. We met once a week for a 3-hour session (4:00–7:00 pm IST), a time slot late enough to avoid conflicts with other classes and to accommodate working professionals. This evening schedule also made it feasible to invite overseas experts for guest talks (e.g., 4–7 pm IST overlaps with morning working hours in Western time zones).
    
    \item \textbf{Semester alignment}: As it is a regular course in each of the institutions, it is offered for a semester. Each institution has their own semester start and end dates and final exam dates. In our case, the phase shift between semester starting dates was only 1 week. We started the course at the earlier of the starting dates and requested the students from other institution to join. The alternative was to start the semester at the later of the starting dates, and give some reading assignments (or some lectures / viewing of online seminars) for the duration till the “course starts”.
    
    \item \textbf{Industry participation}: Most institutes allow guest lectures from industry. However, the challenge lies in enrollment of students from industry. In our case, Both institutions allowed external “casual students” to register for courses (for a fee, yielding an official transcript). To encourage industry participation, we also offered the possibility of auditing the course to industry professionals free of charge (there is no fee and there is no official transcript but we provided letters to those who “completed” the course). 
     
     \item \textbf{Course Assessment}: The outcomes of a research based course can vary in terms of their significance, originality, and methodology. The course has multiple components and multiple faculty. Assessment in such a scenario is a challenge. To avoid subjectivity, we developed common grading rubrics for assignments, presentations, and projects to ensure consistency and objectivity. Instructors of the specific institute were tasked with grading the students from their own institute using the common grading rubrics as guideline. This allowed each institution to follow its grading policies while preserving consistency in standards across institutions.
\end{itemize}

We also note that running a shared course required a common online platform for materials and interaction. We used Google Classroom (GC) as the joint Learning Management System (LMS) so that all students (from both universities and industry) could access the same course space. Lectures were conducted live via video conferencing tools (Google Meet or Zoom) to enable real-time interaction.

\section{Experience with Offering the Course}

In this section, we share further information on conducting such a course, experience with it, and the feedback we received. 

\subsection{Course Set up and Enrollment}

The course was offered jointly by two institutions in two different cities in India.  The authors of the paper were the instructors from the two institutions. The instructors first discussed and agreed on a 1-page course description containing the objectives, topics and structure of the course. Using this description, an elective course was floated in both the institutions (was floated as a "Topics in Software Engineering" course with the specific sub-topic, AI in Software Engineering).  

An illustrative list of topics was also given. The list of topics were on how to use AI techniques (including machine learning) on different aspects of software engineering like requirements engineering, code refactoring, testing, bug fixing, etc. After coordination between instructors on scheduling and content, each university opened enrollment to graduate and senior undergraduate students on the research track. As expected for an advanced niche course, enrollment was modest: 5 students from one institution, and 7 from the other. Two additional PhD students from a third university participated informally (their university did not allow external course credits). The instructors used their network to let industry people know about the possibility of enrolling in the course. A total of 15 working professionals from industry registered in the course via the audit mode. In total, the class had 27 participants, roughly half of them from industry. 

\subsection{Teaching / Lectures}

We adopted a hybrid lecture model leveraging both instructor expertise and guest speakers. Each instructor led sessions on foundational material and selected research topics within their expertise. In addition, we invited external experts (from companies such as TCS, Accenture, IBM, Microsoft, Google, etc.) and academics from other institutes to deliver guest lectures on specialized topics. To ensure students had the necessary background, the course began with tutorial-style sessions on key AI techniques (one session each on supervised learning, unsupervised learning, and recommender systems)

After these foundations, each weekly session focused on a specific “AI for SE” topic. We introduced the Software Engineering context of the problem and then discussed how AI/ML techniques are applied to that domain. Some of the topics are listed below:

\begin{itemize}
    \item Use of NLP/IR in Requirements Engineering
    \item Software effort estimation using AI
    \item Sankie: An AI for DevOps Case Study
    \item AI for Code Generation - large language models, neuro-symbolic reasoning
    \item Advances in Code Summarization
    \item Learning-based Assistant for Data Migration of Enterprise Information Systems
    \item AI for API testing
    \item Auto Bug Repair
    \item Influence of AI on Programming Languages
    \item Code clone detection
    \item Process Mining of Software Repositories
    \item Software Analytics
\end{itemize}

As the course offering predated the modern generative AI, topics like large language models for code were not covered. Overall, about a dozen of the weekly sessions were conducted by invited guest lecturers, complementing the classes taught by the instructors.

\subsection{Student Research Projects}

Students were required to do a research project in the course. They had to select a topic, write a state of the art report on it, present a proposal for research, and finally write a short report and present the results of the research. All the students and many industry professionals did the project and presented the results in the class. Most groups did small projects with some of them applying some ML technique for some problem. In some of the projects they obtained data from GitHub, StackOverflow, etc.  Some of the titles of final reports were: 

\begin{itemize}
	\item Analysis of Architecture Cues in Requirements
	\item Efficient Software Bug Triaging using Machine Learning and NLP Techniques
	\item Generating Pull Request Descriptions using AI
	\item Source Code Vulnerability Analysis
	\item Generating Commit Messages for Summarizing Code
	\item Bug Localization using Structural Information of Source Code
	\item EcoTagger - Tagging Energy Sensitive GitHub Issues
	\item Software Defect Prediction using CNN
\end{itemize}

Most had developed some approach and applied it to some public data set. The presentations indicated that most had obtained some results in these somewhat small research problems. The quality of work of many of these was publishable. However, it was left up to the learners to followup and submit for publication. To give an idea of the type of research work done in these projects, brief descriptions (taken from the abstracts of the submissions) of three projects are given below - one from each of the two academic institutions and one from industry participants.

\begin{itemize}
\item \textbf{Source Code Vulnerability Analysis} (Project by industry participants): Detecting vulnerabilities in the source code of software systems is imperative. This project explores the development of software vulnerability detection using machine learning, particularly convolutional neural networks (CNN) and recurrent neural networks (RNN) to interpret lexed C/C++ source code. The dataset used contains millions of open-source functions that were compiled and labeled from static analyzers that reveal potential exploits. Our experiments show that ML based approach is promising and with better tuning they can help drastically reduce the effort in detecting vulnerabilities.
\item \textbf{Generating Pull Request Descriptions using AI} (by a group of students from one Institution):  Writing a good pull request description is important for readers to understand the changes made, classify and assign the PR to a reviewer, etc. However, in a open source data set it was found that appx 34\% PRs had an empty pull request description, though most had few short but descriptive comments and commit messages. We  propose to generate pull request descriptions from these assets by employing abstractive summarisation approaches. Our approach achieves ROUGE-1, ROUGE-2 and ROUGE-L of 35.32, 21.42 and 36.09 respectively.
\item \textbf{Software Defect Prediction using CNN} (By a group of students from the second institution): Traditional defect prediction techniques often fail to capture the semantic context of the code base, which degrades the performance of the static analysis tools. In this project, we propose to employ a deep learning model for defect prediction, which can learn the semantic representations of programs automatically. We use a convolutional neural network model to automatically learn semantic features using token vectors extracted from the source code abstract syntax trees. Experimental results show that it can improve the effectiveness of defect prediction, sometimes significantly. 
\end{itemize}

\subsection{Feedback and Evaluation}
The overall feedback from participants was positive. Academically, the course succeeded in exposing students to state-of-the-art AI applications in SE. In fact, some student projects led to tangible outcomes beyond the class: one student chose her Master’s thesis topic based on her course project, and another student’s project results were later published in a conference.

At semester end, we collected anonymous feedback via questionnaires. Students rated how well the course met its learning outcomes with an average of 4.4 out of 5 (individual outcome ratings ranged from 4.0 to 4.8). They also gave high marks on quality of instruction and content. In open-ended responses, students particularly appreciated the industry guest lectures (one wrote “The guest lectures were good; provided a lot of insights”). 

On what changes they would like to see to improve learning, one reply was: "Even if it was a research-level course, more clearly defined objectives for the course would have helped. Also, announcing the topic of the lecture about at least 5 days in advance would have also given us some time to review. A more thorough feedback loop during project evaluation in initial phases."

For getting industry feedback, we devised a separate form and requested the attendees to fill it (anonymously). In the feedback we also collected some information about their background and motivation. About 6 participants filled the form. About half the participants had 2-5 yrs experience, and the other half had 10+ yrs experience. About one-third of the attendees self-reported attending most of the sessions, and about 40\% self-reported attending one-third or less sessions. Regarding the reason for joining the course, the dominant reason was: that they were working in that area.

On how valuable are such advanced/research-level courses in Software Engineering for professionals who are working on advanced projects, about 57\% found it very useful, and 28\% found it moderately useful, rest did not find it useful. There were multiple suggestions regarding what else could have been done in the course to make it better. A common theme that evolved was to not just to focus on concepts and approaches but rather on its applicability in industry. This feedback indicates that while professionals value advanced topics, aligning content with practice (e.g., case studies or implementation examples) could improve engagement.

\section{Lessons Learned}
Our experience yielded several insights about the multi-institution, industry-partnered course model. Some aspects of our approach worked very well and we recommend them for similar offerings, while other aspects were less successful and would benefit from adjustments in future iterations.

\subsection{What worked well}

\begin{itemize}
	\item \textbf{Independent Electives with “Guest” Lectures}: We offered the class as separate electives at each university, counting external instructors’ sessions as guest lectures. This workaround allows for offering of such courses within the existing framework and with minimal approvals, if any, from the academic committees or functionaries. 
	\item \textbf{Evening Schedule}: Holding the class in the late afternoon/evening (once a week, 3-hour block) avoided conflicts with other courses and room schedules. It also made the course accessible to industry professionals (who could attend after the workday) and enabled us to include guest speakers from overseas (due to overlapping time zones). This scheduling was instrumental in achieving strong industry participation. 
	\item \textbf{Audit option for industry}: Allowing professionals to audit the course (attend without formal enrollment or fees) significantly lowered barriers to industry involvement. Although industry had Leaning and Development departments that can formalize such registrations, it usually involves a cumbersome process.  Many industry participants likely would not have gone through a lengthy registration process or paid tuition for a single course that's focused on research; the audit mechanism made it easy for them to join and learn, contributing to the diverse participant mix.
\end{itemize}

\subsection{Challenges and Areas of Improvement}
\begin{itemize}
	\item \textbf{Project Guidance and Team Mixing}: We found that students wanted more structured guidance for projects. In future offerings, providing a detailed project guideline and timeline (with interim milestones) could improve project outcomes. Additionally, most project teams were homogenous (members from the same institution or company). If cross-institution or student–professional collaboration is a goal, instructors should introduce guidelines or incentives to form mixed teams. This would give participants experience in distributed collaboration, akin to global software engineering projects.
	\item \textbf{Up-front Course Planning}: Due to time constraints, we did not finalize the entire list of topics before the semester began; instead, we maintained a broad scope and confirmed specific topics as the course progressed. A clear lesson is that more upfront planning would be beneficial. Future multi-institution courses should allot a longer lead time for instructors to jointly design a detailed syllabus and schedule before the course starts. This would also address student feedback about clearly defined objectives and allow everyone (students and guest speakers alike) to prepare for each session well in advance. Further this will also allow for a proper sequencing of the guest lectures.
	\item \textbf{Sustaining Industry Engagement}: Although initial enrollment from industry was high, many professionals only attended the lectures most relevant to them. For a better learning community, it would be ideal to have stronger commitment from all participants. In the future, ensuring industry engagement might require additional measures. For example, having the learners’ employers support their participation (through manager approval or allotted training hours) could improve accountability. Adjusting course content to include tangible takeaways for practice (as suggested in feedback) might also encourage professionals to stay engaged throughout the course. Some of the projects required using datasets for their experiment. Instead of public datasets, industry personnel can probably use their own datasets with appropriate levels of data privacy built in. Such experimental work can potentially help with their regular day jobs too.
    \item \textbf{Enhanced Student Motivation and Outcomes}: We observed that students took the course very seriously, likely due to the unique opportunity to interact with experts outside their home institution and industry peers. This was reflected in the quality of their projects, some of which continued beyond the semester (leading to thesis work and publications, as noted). A lesson here is that creating a platform with broader exposure can inspire students to produce higher-quality work and pursue further research.
    \item \textbf{Academic–Industry Knowledge Exchange}: Having industry professionals and academics learn together proved mutually beneficial. Students were exposed to real-world perspectives and could gauge the practical relevance of research ideas, while the professionals updated their knowledge on cutting-edge academic research. Class discussions often benefited from practitioners bringing up industry use cases. 
\end{itemize}

\section{Conclusion}
In many academic institutions where the size of PhD programs is modest, offering research level courses is extremely challenging. Covid has forced institutions to set up infrastructure necessary to move to online mode of teaching and learning. Students, faculty, and industry professionals are now used to this mode and find it acceptable to continue in cases where in-person sessions are not possible. This infrastructure can be leveraged to offer multi-institutional courses for advanced subjects where individual institutions may not have the capability to offer and/or may not have enough research students to enroll. In this paper we have described an experiment in teaching a course “AI in Software Engineering” jointly between two institutions, and with the industry participating actively. 

Our experience shows that there is a good interest in industry participants to attend such a course, and in industry experts to deliver lectures. The experience suggests that both university students and participants from industry find such a course useful and effective.  The course also helped faculty in developing a deeper understanding of the area, as well as of each other's work, which can pave the way for collaboration in this new area.

There are many other potential advantages of this approach of offering research-level courses across institutions and with participation from industry. For example:

\begin{itemize}
	\item Can enhance collaboration between institutions, as active researchers from these organizations are co-teaching a course on a focused topic.
	\item Can enhance industry-academia linkages.
	\item Can build a small virtual research community in the area, which can be leveraged by faculty from the participating institutions later in different ways (e.g. collaborative projects). 
	\item Can help faculty in smaller  institutions in improving their research. In a place like India, beyond the top institutions, very little research happens in academic institutions. However, some faculty from these institutions are motivated to do research but they have no peers in their institution to help and guide. A faculty member from such an institution who wishes to do research in the area can also enroll in such a course and use it to build their knowledge base in the area, and get some experience in conducting some research.
	\item Can facilitate international collaboration. The institutions participating in this approach can be from different countries. Nowadays people are used to scheduling meetings across time zones. This can help international collaboration as well as research capability building in countries where this may be needed.
\end{itemize}

Overall, we feel that for universities with smaller research programs, such collaboration can help strengthen their research and improve the preparation of their PhD students, and alleviate some of  the disadvantages that come due to lack of ability to offer research level courses to the (few) research students that such universities have. Other advantages can also accrue with such courses. We feel that if other institutions also experiment with this approach, the collective experience can lead to formulating suitable guidelines for institutions for facilitating offering of such courses.


\end{document}